\documentclass[11pt,letter]{article}
\pdfoutput=1
\usepackage{jheppub}
\makeatletter
\def\@fpheader{\bigskip\relax}
\makeatother

\usepackage{amsfonts,amssymb}
\usepackage[mathscr]{eucal}
\usepackage{amsmath}
\usepackage{epsfig,graphicx}
\usepackage{bbold}
\usepackage{color}
\usepackage{datetime}
\usepackage[utf8]{inputenc}

\DeclareMathOperator\arctanh{arctanh}

\newcommand{\be}{\begin{equation}}
\newcommand{\ee}{\end{equation}}
\newcommand{\ba}{\begin{eqnarray}}
\newcommand{\ea}{\end{eqnarray}}
\newcommand{\beg}{\begin{gather*}}
\newcommand{\eng}{\end{gather*}}

\newcommand{\const}{{\rm const}}

\newcommand{\hh}{,\hspace{0.6cm}}
\newcommand{\hhh}{,\hspace{0.2cm}}

\newcommand{\eq}[1]{(\ref{#1})}

\newcommand{\m}{{\scriptscriptstyle -}}
\newcommand{\p}{{\scriptscriptstyle +}}

\newcommand{\ins}[1]{{\mbox{\tiny #1}}}

\def\XXint#1#2#3{{\setbox0=\hbox{$#1{#2#3}{\int}$ }
\vcenter{\hbox{$#2#3$ }}\kern-.6\wd0}}

\title{\boldmath On the Liouville 2D dilaton gravity models with sinh-Gordon matter}

\author[]{Valeri P. Frolov}
\author[]{and Andrei Zelnikov}

\affiliation[]{Theoretical Physics Institute, Department of Physics,\\
University of Alberta, Edmonton, Alberta T6G 2E1, Canada}

\emailAdd{vfrolov@ualberta.ca}
\emailAdd{zelnikov@ualberta.ca}

\abstract{
We study $1+1$ dimensional dilaton gravity models which take into account backreaction of the $\sinh$-Gordon matter field. We found a wide class of exact solutions which generalizes black hole solutions of the Jackiw-Teitelboim gravity model and its hyperbolic deformation.
}
\keywords{Dilaton gravity, Liouville model, Black Holes}

\begin{document}
\maketitle


\section{Introduction}

The $1+1$ dimensional dilaton gravity models, that on the one hand are exactly solvable and on the other hand have quite a rich structure, are useful tools to study many different aspects of classical and quantum gravity (for comprehensive reviews see \cite{Grumiller:2002nm,Fabbri:2005mw,Nakayama:2004vk}). These models naturally appear after dimensional reduction of spherically symmetric gravity and its generalizations and provide an ideal testing ground for the study of black hole quantum mechanics, thermodynamics, and such deep issues in quantum gravity as the endpoint of Hawking radiation and the nature of black hole entropy. Special 2D dilaton gravity models were extensively used in the study of classical and quantum properties of Liouville black holes \cite{Mann:1993rf}, the exact string and CGHS black holes \cite{Grumiller:2005sq,Dijkgraaf:1991ba,Witten:1991yr,Mandal:1991tz,Callan:1992rs}. The Jackiw-Teitelboim model \cite{Jackiw:1984je,Teitelboim:1983ux} is one of the simplest 2D dilaton gravity theories which contains most of the desirable features. Recently it has been used to  study  backreaction effects in asymptotically $\mathrm{AdS_2}$ spacetimes in the context of holography \cite{Almheiri:2014cka}.  Its deformation has been found \cite{Kyono:2017jtc,Kyono:2017pxs} using Yang-Baxter deformation technique, so that the quadratic potential is replaced by a hyperbolic function of the dilaton field.

In this paper we use a simpler method to derive the hyperbolic deformation \cite{Kyono:2017jtc,Kyono:2017pxs} of the Jackiw-Teitelboim model and further generalize it to include  $\sinh$-Gordon type matter field. The class of the proposed dilaton gravity models is still exactly solvable for a variety of choices of matter and dilaton distributions.


\section{The model}

We begin by considering a model of a two-dimensional dilaton gravity which interacts with a matter field $\chi$ of the $\sinh$-Gordon type and a free massless scalar field $f$.
The action functional of the system we are interested in reads
\be\begin{split}\label{S0}
&S~=S[g_{\mu\nu},\phi,\chi,f]=S_\phi+S_\chi+S_f,\\
&S_\phi=-{1\over 4\pi}\int d^2x\,\sqrt{-g}\Big[
Q\phi R+(1-2bQ)( \nabla\phi)^2-\lambda\, e^{-2b\phi}
\Big],\\
&S_\chi=-{1\over 4\pi}\int d^2x\,\sqrt{-g}\Big[
( \nabla\chi)^2+m\,\cosh(2b\chi)
\Big],\\
&S_f=-{1\over 8\pi}\int d^2x\,\sqrt{-g}\,(\nabla f)^2.
\end{split}\ee
The action $S_\phi$ is the action of the Liouville dilaton gravity. On a flat background the action $S_\chi$ describes the matter field $\chi$ which obeys the well known $\mathbf{\sinh}$-Gordon equation.  However, in our case the geometry is curved. The other field $f$ is a free conformal matter useful, e.g., to study the solutions involving collapsing null shells.

We consider the model with positive parameters $\lambda$ and $m$. By shifting the dilaton field as $\phi\to\phi+\const$, one can make the coefficient $\lambda$ in front of the Liouville potential to become an arbitrary (positive) constant. This is possible because in two dimensions $\int R$ is a topological invariant and this field redefinition does not affect the field equations. Later on we will put $\lambda=m$.

Using conformal transformations of the metric, the action \eq{S0} can be rewritten is several different but equivalent  forms. In the literature it is often used the representation when the kinetic term of the dilaton field $\phi$ cancels. This choice is given by the following metric redefinition
\be\label{barg}
g_{\mu\nu}=e^{\big(2b-{1\over Q}\big)\phi}\hat{g}_{\mu\nu}.
\ee
When expressed in terms of the $\hat{g}_{\mu\nu}$ metric the action functional \eq{S0} becomes
\be\begin{split}\label{S2}
\hat{S}&=-{1\over 4\pi}\int d^2x\,\sqrt{-\hat{g}}\Big[
Q\phi \hat{R}
+( \hat{\nabla}\chi)^2
+m\,e^{\big(2b-{1\over Q}\big)\phi}\cosh(2b\chi)-m\, e^{-{1\over Q}\phi}+{1\over 2}(\hat{\nabla} f)^2
\Big].
\end{split}\ee

From the computational point of view the following conformal transformation happens to be more convenient
\be\begin{split}\label{g}
&g_{\mu\nu}=e^{2b\phi}\tilde{g}_{\mu\nu}\hh
\sqrt{-g}=e^{2b\phi}\sqrt{-\tilde{g}}\hh
R=e^{-2b\phi}[\tilde{R}-2b\,\tilde{\Box}\phi].
\end{split}\ee
Then the action \eq{S0} takes the form
\be\begin{split}\label{S1}
\tilde{S}&=-{1\over 4\pi}\int d^2x\,\sqrt{-\tilde{g}}\Big[
Q\phi \tilde{R}+( \tilde{\nabla}\phi)^2+( \tilde{\nabla}\chi)^2
+m\,e^{2b\phi}\cosh(2b\chi)-m+{1\over 2}(\tilde{\nabla} f)^2
\Big],
\end{split}\ee

It is also useful to introduce two other field variables
\be
\omega_1=\phi+\chi  \hh
\omega_2=\phi-\chi.
\ee
In terms of these fields the action \eq{S1} becomes a sum of two decoupled Liouville-type actions\footnote{This representation is similar to that of \cite{Kyono:2017pxs}, where the difference of Liouville actions was considered.} and the action of a free scalar field
\be\label{S3}
\tilde{S}=S_1+S_2+S_f,
\ee
where
\be\begin{split}\label{Somega}
S_{k}=&-{1\over 8\pi}\int d^2x\,\sqrt{-\tilde{g}}\Big[
\big(Q\omega_k \tilde{R}
+( \tilde{\nabla}\omega_k)^2+m\,e^{2b\omega_k}-m\big)
\Big],
\end{split}\ee
and $ k=(1,2) $. These Lagrangians differ from the standard Liouville field theory only in the extra ``cosmological" term $-m$.
Variation of the action \eq{S3} over the fields $\omega_k$ (for $k=(1,2)$) and $f$ gives the field equations
\be\begin{split}\label{fieldEQ}
&\tilde{\Box}\omega_k-{Q\over 2}\tilde{R}-bm\,e^{2b\omega_k}=0,\\
&\tilde{\Box}f=0.
\end{split}\ee
Variation of the action \eq{S3} over the metric leads to the equations
\be\label{Tmn0}
\tilde{T}^{\mu\nu}=T^{\mu\nu}_1+T^{\mu\nu}_2+T^{\mu\nu}_f=0,
\ee
where
\be
\tilde{T}^{\mu\nu}={2\over\sqrt{-\tilde{g}}}{\delta \tilde{S}\over\delta \tilde{g}_{\mu\nu}} \hh T^{\mu\nu}_k={2\over\sqrt{-\tilde{g}}}{\delta S_k\over\delta \tilde{g}_{\mu\nu}}.
\ee
Here
\be\begin{split}\label{Tmn}
T^{\mu\nu}&=-{1\over 8\pi}\Big[2\big(Q\omega{}^{;\mu\nu}-\omega{}^{;\mu}\omega{}^{;\nu}\big)
+\tilde{g}^{\mu\nu}\big(-2Q\omega{}^{;\alpha}{}_{\alpha}+\omega{}^{;\alpha}\omega{}_{;\alpha}+m(e^{2b\omega}-1)\big)
\Big]
\end{split}\ee
for $\omega=(\omega_1,\omega_2)$ correspondingly and
\be
T^{\mu\nu}_f={1\over 4\pi}\Big[f^{;\mu}f^{;\nu}
-{1\over 2}\tilde{g}^{\mu\nu}f^{;\alpha}f_{;\alpha}
\Big].
\ee

In the conformal gauge we have
\be\label{conf}
d\tilde{s}^2=-2 e^{2\rho}dx^\p dx^\m \hh x^\p=t+z\hh x^\m=t-z,
\ee
\be\begin{split}
&\tilde{g}_{\scriptscriptstyle ++}=\tilde{g}_{\scriptscriptstyle --}=0\hh \tilde{g}_{\scriptscriptstyle +-}=\tilde{g}_{\scriptscriptstyle -+}=-e^{2\rho},
\\
&\tilde{g}^{\scriptscriptstyle ++}=\tilde{g}^{\scriptscriptstyle --}=0\hh \tilde{g}^{\scriptscriptstyle +-}=\tilde{g}^{\scriptscriptstyle -+}=-e^{-2\rho}  \hh
\sqrt{-\tilde{g}}=e^{2\rho}.
\end{split}\ee

Let us denote
\be\begin{split}
&~\,\omega_\p=\partial_\p\omega\hh~~~~~
\omega_\m=\partial_\m\omega
,\\
&\omega_{\p\p}=\partial_\p\partial_\p\omega \hh
\omega_{\m\m}=\partial_\m\partial_\m\omega,\\
&\omega_{\p\m}=\partial_\m\partial_\p\omega \hh
\omega_{\m\p}=\partial_\p\partial_\m\omega ,
\end{split}\ee
and introduce similar objects for partial derivatives of $\rho$, $f$, and $\sigma$. Then
\be\label{tildeR}
\tilde{R}=4 e^{-2\rho}\rho_{\p\m}\hh
\tilde{\Box}\rho=-2e^{-2\rho}\rho_{\p\m},
\ee
\be
\tilde{\Box}\omega=-2e^{-2\rho}\omega_{\p\m}
\hh
\tilde{\Box}f=-2e^{-2\rho}f_{\p\m}.
\ee
The components of the stress-energy $T_{\mu\nu}$ (see \eq{Tmn}) take the form
\be\begin{split}\label{Tab1}
&T_{\p\p}={1\over 4\pi}\Big[-Q\omega_{\p\p}+\omega_{\p}\omega_{\p}+2Q\omega_{\p}\rho_{\p}\Big],\\
&T_{\m\m}={1\over 4\pi}\Big[-Q\omega_{\m\m}+\omega_{\m}\omega_{\m}+2Q\omega_{\m}\rho_{\m}\Big],\\
&T_{\p\m}=T_{\m\p}={1\over 8\pi}\Big[2Q\omega_{\p\m}+m\,e^{2\rho}\big(e^{2b\omega}-1\big)\Big].
\end{split}\ee

Field equations \eq{fieldEQ} become
\be\begin{split}\label{fieldsEQ1}
&\omega_k{}_{\p\m}+Q\rho_{\p\m}+{bm\over 2}e^{2(\rho+b\omega_k)}=0,\\
&f_{\p\m}=0.
\end{split}\ee
Variation of the action \eq{S3} over the metric leads to \eq{Tmn0}. When written explicitly in components these equations take the form
\be\begin{split}
2Q(\omega_1{}_{\p\m}&+\omega_2{}_{\p\m})
+m\,e^{2\rho}\big(e^{2b\omega_1}+e^{2b\omega_2}-2\big)=0,\label{Tab2a}
\end{split}\ee
\be\begin{split}
\omega_1{}_{\p}\omega_1{}_{\p}&-Q\omega_1{}_{\p\p}+\omega_2{}_{\p}\omega_2{}_{\p}-Q\omega_2{}_{\p\p}
+2Q(\omega_1{}_{\p}+\omega_2{}_{\p})\rho_{\p}+f_{\p}f_{\p}=0,\label{Tab2b}
\end{split}\ee
\be\begin{split}
\omega_1{}_{\m}\omega_1{}_{\m}&-Q\omega_1{}_{\m\m}+\omega_2{}_{\m}\omega_2{}_{\m}-Q\omega_2{}_{\m\m}
+2Q(\omega_1{}_{\m}+\omega_2{}_{\m})\rho_{\m}+f_{\m}f_{\m}=0\label{Tab2c}.
\end{split}\ee

From \eq{fieldsEQ1} and \eq{Tab2a} we get
\be\label{bQ}
(1-bQ)(\omega_1{}_{\p\m}+\omega_2{}_{\p\m})+2Q\rho{}_{\p\m}+m be^{2\rho}=0.
\ee
For a specific value of the constant $Q=1/b$ the equation for the conformal factor decouples from the matter equations. This case is of a particular interest for us, because the classical field equations can be solved exactly. In the next sections we fix $Q=1/b$.


\section{Case $Q=1/b$.}

In a special case, when $Q=1/b$, \eq{bQ} reduces to the Liouville equation
\be\label{rho0}
\rho_{\p\m}+{m b^2\over 2}e^{2\rho}=0.
\ee
Taking into account \eq{tildeR}, one can see that the solution for the metric $\tilde{g}_{\mu\nu}$ describes the spacetime of a constant curvature
\be
\tilde{R}=-2m b^2.
\ee
The general solution of the Liouville equation is well known and reads
\be\label{rho}
e^{2\rho}={2\over m b^2}\,{\partial_\p Y^\p\partial_\m Y^\m\over(Y^\p-Y^\m)^2},
\ee
where
\be
Y^\p=Y^\p(x^\p)\hh
Y^\m=Y^\m(x^\m)
\ee
are arbitrary functions of the advanced and retarded null coordinates.

The other field equations are
\be\begin{split}\label{fieldsEQ3}
&(b\,\omega_k+\rho)_{\p\m}+{m b^2\over 2}e^{2(b\,\omega_k+\rho)}=0.
\end{split}\ee
It is convenient to introduce new fields
\be
\sigma_k=b\,\omega_k+\rho.
\ee
In terms of these variables \eq{fieldsEQ3} take the form
\be\begin{split}\label{fieldsEQ4}
&\sigma_k{}_{\p\m}+{m b^2\over 2}e^{2\sigma_k}=0.
\end{split}\ee
Thus, in the conformal gauge the field equations for the metric and two $\sigma$-fields reduce to three identical Liouville equations. The constraint equations \eq{Tab2b},\eq{Tab2c} mean that total fluxes $\tilde{T}_{\p\p}=\tilde{T}_{\m\m}=0$ and in terms of $\sigma$ fields read
\be\begin{split}
\sigma_1{}_{\p\p}-(\sigma_1{}_{\p})^2 &+\sigma_2{}_{\p\p}-(\sigma_2{}_{\p})^2
=2\big[\rho_{\p\p}-(\rho_{\p})^2\big]+b^2(f_\p)^2,           \label{Tab3b}
\end{split}\ee
\be\begin{split}
\sigma_1{}_{\m\m}-(\sigma_1{}_{\m})^2 &+\sigma_2{}_{\m\m}-(\sigma_2{}_{\m})^2
=2\big[\rho_{\m\m}-(\rho_{\m})^2\big]+b^2(f_\m)^2.        \label{Tab3c}
\end{split}\ee
The matter field $f$ satisfies the equation
\be
f_{\p\m}=0.
\ee
The general solution of this equation is described by two arbitrary functions $V$ and $U$
\be\label{UV}
f=V(x^\p)+U(x^\m).
\ee

The solutions of the \eq{fieldsEQ4}, which have the form of the Liouville equation, are
\be\label{sigma12}
e^{2\sigma_k}={2\over mb^2}\,{\partial_\p X^\p_k\partial_\m X^\m_k\over(X^\p_k-X^\m_k)^2}\hh k=(1,2),
\ee
where
\be
X^\pm_k=X^\pm_k(x^\pm)
\ee
are arbitrary functions of the advanced and retarded null coordinates.
For the fields $\omega_k$ it leads to
\be\label{omega12}
e^{2b\omega_k}={\partial_\p X^\p_k\partial_\m X^\m_k\over(X^\p_k-X^\m_k)^2}\,{(Y^\p-Y^\m)^2\over \partial_\p Y^\p\partial_\m Y^\m}.
\ee
Using this solution and \eq{rho} one can find the original metric \eq{g}, the dilaton field $\phi$ and the matter fields $\chi$.
\be\label{phi}
e^{2b\phi}={\sqrt{\partial_\p X^\p_1\partial_\m X^\m_1~ \partial_\p X^\p_2\partial_\m X^\m_2}\over|X^\p_1-X^\m_1||X^\p_2-X^\m_2|}\,{(Y^\p-Y^\m)^2\over \partial_\p Y^\p\partial_\m Y^\m},
\ee
\be\label{chi}
e^{2b\chi}=\sqrt{{\partial_\p X^\p_1\partial_\m X^\m_1\over(X^\p_1-X^\m_1)^2}}\sqrt{{(X^\p_2-X^\m_2)^2\over \partial_\p X^\p_2\partial_\m X^\m_2}}.
\ee

The result of substitution of the solutions \eq{sigma12} to \eq{Tab3b}-\eq{Tab3c} can be written in the form
\be\begin{split}\label{XXY}
&\{X_1^\p,x^\p\}+\{X_2^\p,x^\p\}=2\{Y^\p,x^\p\}+2b^2(f_\p)^2 ,\\
&\{X_1^\m,x^\m\}+\{X_2^\m,x^\m\}=2\{Y^\m,x^\m\}+2b^2(f_\m)^2.
\end{split}\ee
Here $\{A,x\}$ denotes the Schwarzian derivative of the function $A(x)$
\be
\{A,x\}\equiv {A'''\over A'}-{3\over 2}\Big({A''\over A'}\Big)^2.
\ee

One can always use a coordinate transformation such that
\be\label{Y}
Y^\p=x^\p  \hhh Y^\m=x^\m  \hhh e^{2\rho}={2\over m b^2}\,{1\over(x^\p-x^\m)^2}.
\ee
In this gauge the metric becomes
\be
d\tilde{s}^2=-{4\over m b^2}\,{dx^\p dx^\m\over (x^\p- x^\m)^2}.
\ee
and, evidently,
\be
\{Y^\p,x^\p\}=\{Y^\m,x^\m\}=0.
\ee
The remaining nontrivial equations \eq{XXY} for the dilaton fields reduce to
\be\begin{split}\label{XX}
&\{X_1^\p,x^\p\}+\{X_2^\p,x^\p\}=2b^2(f_\p)^2 ,\\
&\{X_1^\m,x^\m\}+\{X_2^\m,x^\m\}=2b^2(f_\m)^2.
\end{split}\ee

Different choices of the functions $X^\pm_k$ and $f^\pm$ correspond to different physical setups of the problem.


\section{Vacuum  solutions.}

Now consider the vacuum solutions of the system \eq{XX}, when the matter field $f$ vanishes.
Let the functions $X^\pm_k$ be related according to the rule
\be
X_1^{\pm}=X^{\pm} \hh X_2^{\pm}={\alpha X^\pm+\beta\over \gamma \,X^\pm+\delta}
\ee
for some arbitrary coefficients $\alpha,\beta,\gamma,\delta$. Substitution of this relation to \eq{sigma12},\eq{omega12} gives
$
\sigma_1=\sigma_2$ and $\phi=\omega_1=\omega_2
$,
that corresponds to
$
\chi=0
$
in the original field variables. Thus this ansatz assumes that the $\sinh$-Gordon matter field $\chi$ also vanishes.

Thus the choice in question describes the system with the action
\be\begin{split}\label{S01}
\tilde{S}=-{1\over 4\pi}\int d^2x\,&\sqrt{-\tilde{g}}\Big[
{1\over b}\phi \tilde{R}+( \tilde{\nabla}\phi)^2
+m\big(e^{2b\phi}-1\big)+{1\over 2}(\tilde{\nabla} f)^2
\Big].
\end{split}\ee
In terms of the metric  $\hat{g}_{\mu\nu}$ (see \eq{barg}) it reads
\be\begin{split}\label{S02}
\hat{S}=-{1\over 4\pi}\int d^2x\,&\sqrt{-\hat{g}}\Big[
{1\over b}\phi \hat{R}
+2m\,\sinh(b\phi)+{1\over 2}(\hat{\nabla} f)^2
\Big].
\end{split}\ee
One can see that the action \eq{S02} exactly reproduces, up to normalization factors,  the hyperbolic deformation (see \cite{Kyono:2017jtc,Kyono:2017pxs}) of the
Jackiw–Teitelboim gravity model.

The vacuum solution assumes that the matter field $f=0$ vanishes. Then because of the properties of the Schwarzian derivative we obtain
\be
\{X_1^\pm,x^\pm\}=\{X_2^\pm,x^\pm\}=\{X^\pm,x^\pm\}.
\ee
The constraint equations \eq{XX} reduce to
\be\label{XX0}
\{X^\pm,x^\pm\}=0.
\ee
Their general solutions are
\be\label{Xpm}
X^\p(x^\p)={\alpha_1 x^\p+\beta_1\over \gamma_1 \,x^\p+\delta_1}\hh \alpha_1\delta_1-\beta_1\gamma_1 >0,
\ee
\be
X^\m(x^\m)={\alpha_2 \,x^\m+\beta_2\,\over \gamma_2 \,x^\m+\delta_2}\hh \alpha_2\delta_2-\beta_2\gamma_2 >0.
\ee
In the conformal gauge \eq{Y} we obtain the vacuum solution for the dilaton
\be\label{rho1}
e^{2\rho}={2\over m b^2}\,{1\over(x^\p-x^\m)^2},
\ee
\be\label{ephi}
e^{2b\phi}={C_1(x^\p-x^\m)^2  \over
C_2 x^\p x^\m+C_3x^\p-C_4x^\m
+C_5},
\ee
where
\be\begin{split}
C_1&=(\alpha_1\delta_1-\beta_1\gamma_1)(\alpha_2\delta_2-\beta_2\gamma_2),\\
C_2&=\alpha_1\gamma_2-\alpha_2\gamma_1\hh
C_3=\alpha_1\delta_2-\beta_2\gamma_1,\\
C_4&=\alpha_2\delta_1-\beta_1\gamma_2\hh \,
C_5=\beta_1\delta_2-\beta_2\delta_1.
\end{split}\ee
By lifting the solution to a higher dimensional spherical spacetime, the dilaton field gets a meaning of some power of the radial coordinate. In higher dimensions, even for non-static spacetimes, one can define the apparent horizon using the condition, that a normal to the constant radius surface becomes null on the apparent horizon. This property boils down to the following condition for the analogue of the apparent horizon in the two dimensional dilaton gravity
\be\label{hor}
(\tilde{\nabla} \phi)^2\big|_\ins{Hor}=0.
\ee
As an example consider a particular choice of parametrization of the solution \eq{Xpm}
\be\label{Xpm0}
X^\p(x^\p)={(1+2ac) x^\p+pc \over c \,x^\p +1},
\ee
\be
X^\m(x^\m)=x^\m.
\ee
where $a,c,p$ are arbitrary constants satisfying the condition $1+2ac>pc^2$.
Then the dilaton field takes the form
\be
e^{2b\phi}={(1+2ac-pc^2)(x^\p-x^\m)^2\over [x^\p-x^\m+c\,(p+2ax^\p-x^\p x^\m)]^2}.
\ee
The horizon equation \eq{hor}  gives
\be
x^\p \big|_\ins{Hor}=a\pm\sqrt{a^2+p}\hh x^\m \big|_\ins{Hor}=a\pm\sqrt{a^2+p}.
\ee
The obtained solution reproduces the hyperbolic deformation \cite{Kyono:2017jtc,Kyono:2017pxs} of the
Jackiw–Teitelboim gravity model. It can be also generalized to a non-static case \cite{Kyono:2017jtc,Kyono:2017pxs} that includes the collapsing shell of null matter $f$.


\section{Non-vacuum solutions.}

Now consider a more general setup of the problem. By choosing a different ansatz for functions $F^{\pm}$
\be
X_1^{\pm}=X^{\pm} \hh X_2^{\pm}=F^{\pm}(X^\pm)
\ee
and using the chain rule of the Schwarzian derivative, one can write the constraint equations \eq{XX} in the form
\be\label{XXF}
\{X^\pm,x^\pm\}+{1\over 2}\Big({\partial X^\pm\over\partial x^\pm}\Big)^2\{F^\pm,X^\pm\}=b^2(f_\pm)^2.
\ee

Let us study a few simple examples of deformations of the Jackiw–Teitelboim dilaton gravity, that correspond to different choices of the solutions for matter field $\chi$.  In this section the matter field $f$  is assumed to vanish.

In the conformal gauge \eq{Y} for any given functions $F^\pm$ we have  (see \eq{phi},\eq{chi})
\ba\label{phi1}
e^{2b\phi}&=&{|\partial_\p X^\p\partial_\m X^\m|\sqrt{|(F^\p)'(F^\m)'|}\over|X^\p-X^\m||F^\p-F^\m|}\,(x^\p-x^\m)^2,
\\
e^{2b\chi}&=&{|F^\p-F^\m|\over|X^\p-X^\m|\sqrt{|(F^\p)'(F^\m)'|}} .\label{chi1}
\ea
Here $F\,'\equiv\partial_X F(X)$ and $X^\pm$ are the solutions of \eq{XXF}.
The constraint equations \eq{XXF} define functions $X^\pm$, and for $f=0$ take the form
\be\label{XXF0}
\{X,x\}+{1\over 2}\Big({\partial X\over\partial x}\Big)^2\{F,X\}=0.
\ee
For any chosen function $F(X)$, this equation is a third-order ordinary differential equation. Therefore its general solution $X(x)$ is parameterized by three arbitrary constants. In some cases the solution is quite simple. Let us consider a few natural choices of the function $F$ that admit exact solution of \eq{XXF0} in terms or elementary functions.

It should be noted that  if one finds the solution $X(x)$ for some function $F(X)$, then,  because of properties of the Schwarzian derivative, exactly the same function $X(x)$ is the solution of the problem with $F\to G$, provided the function $G$ is fractional linear transformation of the function $F$
\be\label{G}
G(X)={c_1F(X)+c_2\over c_3F(X)+c_4}\hh  \{G,X\}=\{F,X\}.
\ee
The dilaton $\phi$ and matter field $\chi$ depend on choice of functions $F^\pm$. Thus, starting from any given solution and  using this property we can generate a whole class of physically different solutions for the dilaton and matter fields.

Among all possibilities we single out three simplest types
\ba\label{ABC}
\mbox{{\bf A}\hskip1cm when~}\{F,X\}&=&\alpha\phantom{1\over 2}\label{AA}\\
\mbox{{\bf B}\hskip1cm when~}\{F,X\}&=&{\alpha\over X^2}\label{BB},\\
\mbox{{\bf C}\hskip1cm when~}\{F,X\}&=& {\alpha\over X^4}\label{CC},
\ea
where $\alpha$ is an arbitrary constant.


\subsection{Case A}

In this case a particular solution of the problem $\{F,X\}=\alpha$ reads
\be
F=
\begin{cases}
\tan(aX),& \mbox{for}~~ \alpha=+2a^2,\\
\tanh(aX),&  \mbox{for}~~ \alpha=-2a^2.
\end{cases}
\ee
Using the property \eq{G} we can derive a most general form of the function $F$ in the case {\bf A}.
\be\label{FA}
F=
\begin{cases}
\displaystyle{c_1\, \tan(aX)+c_2\over c_3\,  \tan(aX)+c_4} &\mbox{for}~~ \alpha=+2a^2,\\
\displaystyle{c_1\, \exp(2aX)+c_2\over c_3\, \exp(2aX)+c_4} &\mbox{for}~~ \alpha=-2a^2.
\end{cases}
\ee
One can see that, e.g., the functions $\tanh(aX)$, $\coth(aX)$, and $\exp(\pm 2aX)$ have the same constant Schwarzian derivative.

The general solution of \eq{XXF0}, after substitution there $\{F,X\}=\alpha$, is characterized by three arbitrary constants $q_1,q_2,q_3$
\be\label{solA}
X=\begin{cases}
q_1+{\sqrt{2}\over a}\arctan[q_2(x+q_3)], & \alpha=+2 a^2, \\
q_1+{\sqrt{2}\over a}\arctanh[q_2(x+q_3)], & \alpha=-2 a^2.
\end{cases}
\ee
In order to obtain $F^\pm$ as explicit functions of the coordinates $x^\pm$, one has to substitute the solution \eq{solA} into the function in question \eq{FA}.


\subsection{Case B}

A particular solution in the case {\bf B} is
\be\label{FB}
F=
\begin{cases}
X^n,& \mbox{for}~~ \alpha={1-n^2\over 2},\\
\ln(X),&  \mbox{for}~~ \alpha={1\over 2}.
\end{cases}
\ee
Using the fractional linear transformation \eq{G} of \eq{FB} one can generate the other functions which fulfill the condition \eq{BB}.

The general solution of the constraint equations \eq{XXF0}
\be\label{solB}
X=\exp\big[
q_1+{2\sqrt{2}\over\sqrt{n^2+1}}\arctanh[q_2(x+q_3)]
\big].
\ee
Note that $F(X)\sim\ln X$ corresponds to $n=0$ case.


\subsection{Case C}

It it easy to find a function, satisfying the condition \eq{CC}
\be
F=
\begin{cases}
\tan{a\over X},& \mbox{for}~~ \alpha=+2a^2,\\
\tanh{a\over X},&  \mbox{for}~~ \alpha=-2a^2.
\end{cases}
\ee
The general form of the function $F$ in the case {\bf C} reads
\be\label{FA}
F=
\begin{cases}
\displaystyle{c_1\, \tan{a\over X}+c_2\over c_3\,  \tan{a\over X}+c_4} &\mbox{for}~~ \alpha=+2a^2,\\
\displaystyle{c_1\, \tanh{a\over X}+c_2\over c_3\, \tanh{a\over X}+c_4} &\mbox{for}~~ \alpha=-2a^2.
\end{cases}
\ee

Substituting  the ansatz $\{F,X\}=\alpha/X^4$  to \eq{XXF0} we obtain the solution
\be\label{solC}
X=\begin{cases}
q_1+{\sqrt{2}\over a}\arctan[q_2(x+q_3)], & \alpha=+2 a^2, \\
q_1+{\sqrt{2}\over a}\arctanh[q_2(x+q_3)], & \alpha=-2 a^2.
\end{cases}
\ee


\subsection{Some other cases}

Note that the solutions \eq{solC} coincide with \eq{solA}. It's not surprising because $\{X^{-1},x\}=\{X,x\}$ and, hence, substitution $X\to X^{-1}$ transforms constraint equation \eq{XXF0} of the case {\bf C} to that of the case {\bf A}. In fact, any fractional linear transformation of the function $X$
\be
X(x)\to{\epsilon X(x)+\zeta \over \gamma \,X(x)+\delta}
\ee
with arbitrary coefficients $\epsilon,\zeta,\gamma,\delta$ does not alter the Schwarzian derivative. Therefore the solution \eq{solA} is also a solution of the problem
\be\label{XA}
\{X,x\}+{1\over 2}\Big({\partial X\over\partial x}\Big)^2 \alpha{(\delta\epsilon-\gamma\zeta)^2\over (\gamma X(x)+\delta)^4}=0
\hh \alpha=\pm 2 a^2.
\ee
It corresponds to
\be
F=
\begin{cases}
\tan{a(\delta\epsilon-\gamma\zeta)\over \gamma(\gamma X+\delta)},& \mbox{for}~~ \alpha=+2a^2,\\
\tanh{a(\delta\epsilon-\gamma\zeta)\over \gamma(\gamma X+\delta)},&  \mbox{for}~~ \alpha=-2a^2
\end{cases}
\ee
with
\be
 \{F,X\}=\alpha{(\delta\epsilon-\gamma\zeta)^2\over (\gamma X(x)+\delta)^4}.
\ee
and also to all fractional linear transformations of this function $F$.


\section{Summary}

We found out the explicit expressions \eq{solA},\eq{solB},\eq{solC} in terms of elementary functions for the solutions of the constraint equations \eq{XXF0} in cases A,B, and C. The same solutions are valid for the choice of any other function that is the linear fractional (M\"obius) transformation of the functions we have considered.  Then one has to substitute these solutions $X^\pm(x^\pm)$ to $F^\pm(X^\pm)$ and,using \eq{phi1}-\eq{chi1}, derive the value of the dilaton $\phi$ and the matter field $\chi$.

The metric $\tilde{g}_{\mu\nu}$ (see \eq{conf}) describes pure $\mathrm{AdS_2}$ spacetime with the constant curvature $\tilde{R}=-2m b^2$. The metrics $g_{\mu\nu}$ and $\hat{g}_{\mu\nu}$  (see \eq{g}-\eq{barg}) describe conformal deformations of the $\mathrm{AdS_2}$ spacetime.

The obtained exact classical solutions typically have both horizons and singularities. There are also other solutions we did not present in this paper. Their properties should be analyzed for every particular choice of the matter fields $\chi$ and $f$. Similar to the cases of Jackiw-Teitelboim  gravity \cite{Almheiri:2014cka} or its hyperbolic deformation \cite{Kyono:2017jtc,Kyono:2017pxs}, one can use the derived solutions to describe collapsing matter problem, holography, thermodynamics, and Hawking radiation effects. We will return to these  interesting questions in future publications.

\section*{Acknowledgments}

The authors thank the Natural Sciences and Engineering Research Council of Canada and the Killam Trust for their financial support.


\providecommand{\href}[2]{#2}\begingroup\raggedright\endgroup



\end{document}